\documentclass[12pt]{article}

\usepackage{graphicx}
\usepackage{amsmath,amssymb}

\newcommand{\ignorar}[1]{}

\title{Seasonal and regional characterization of horizontal stirring in the global ocean}

\author{Ismael Hern\'andez-Carrasco,$^{1}$ Crist\'obal L\'opez,$^{1\ast}$\\
Emilio Hern\'andez-Garc\'ia$^{1}$ and Antonio Turiel,$^{2}$ \\
\\
\normalsize{$^{1}$Instituto de F\'isica Interdisciplinar y Sistemas Complejos (CSIC-UIB)}\\
\normalsize{07122 Palma de Mallorca, Spain}\\
\normalsize{$^{2}$Institut de Ci\`encies del Mar, CSIC}\\
\normalsize{Passeig Mar\'{\i}tim de la Barceloneta 37-49, 08003
  Barcelona, Spain}\\
\\
\\
\normalsize{$^\ast$To whom correspondence should be addressed; E-mail: clopez@ifisc.uib.es}
}

\date{} 

\begin{document}

\baselineskip24pt

\maketitle

\newpage

\begin{abstract}

Recent work on Lagrangian descriptors has shown that Lyapunov
Exponents can be applied to observed or simulated data to
characterize the horizontal stirring and transport properties
of the oceanic flow. However, a more detailed analysis of
regional dependence and seasonal variability was still lacking.
In this paper, we analyze the near-surface velocity field
obtained from the {\it Ocean general circulation model For the
Earth Simulator} (OFES) using Finite-Size Lyapunov Exponents
(FSLE). We have characterized regional and seasonal
variability. Our results show that horizontal stirring, as
measured by FSLEs, is seasonally-varying, with maximum values
in Summer time. FSLEs also strongly vary depending on the
region: we have first characterized the stirring properties of
Northern and Southern Hemispheres, then the main oceanic basins
and currents. We have finally studied the relation between
averages of FSLE and some Eulerian descriptors such as Eddy
Kinetic Energy (EKE) and vorticity ($\omega$) over the
different regions.

\end{abstract}

\maketitle

%
%

%


%
%

\section{Introduction}

A detailed knowledge of the transport, dispersion, stirring and
mixing mechanisms of water masses across the global ocean is of
crucial interest to fully understand, for example, heat and
tracer budgets, or the role of oceans in climate regulation.
There has been a recent strong activity in the study of these
processes from a Lagrangian perspective. Some works have
addressed the {\it global} variability of them using
finite-time Lyapunov exponents (FTLEs) computed from currents
derived from satellite altimetry
\cite{BeronVera2008,Waugh2008}. These studies quantify
stirring intensity, and identify mesoscale eddies and other
Lagrangian Coherent Structures (LCSs). Furthermore, previous
works \cite{Waugh2006} pointed out relationships between
Lagrangian and Eulerian quantifiers of stirring/mixing activity
(FTLEs and Eddy Kinetic Energy (EKE) or mean strain rate).

Having in mind the implications for the distribution of
biogeochemical tracers, our goal is to extend the previous
works to provide detailed seasonal analysis and a comparative
study between different ocean regions and different scales:
Earth's hemispheres, ocean basins, and boundary currents. To
this end we use finite-size Lyapunov exponents (FSLEs). These
quantities are related to FTLEs since they also compute
stretching and contraction time scales for transport, but they
depend on explicit spatial scales which are simple to specify
and to interpret in oceanographic contexts \cite{dOvidio2004,
dOvidio2009, HernandezCarrasco2011, TewKai2009}. In particular
we will focus on the impact on transport of mesoscale
processes, for which characteristic spatial scales as a
function of latitude are well known. We are also interested in
checking the existence of relationships between Lagrangian
measures of horizontal stirring intensity, as given by averages
of finite-size Lyapunov exponents (FSLE), and other dynamic,
Eulerian quantities, such as EKE or vorticity. Such a
functional relation does not need to hold in general, but may
be present when there is a connection between the mechanisms
giving rise to mesoscale turbulence (probably, baroclinic
instability) and horizontal stirring.

The paper is organized as follows. In Section \ref{Sec:datamet}
we describe the data and tools used in this study. In section
\ref{Sec:results} we first present the geographical and
seasonal characterization of the horizontal stirring, and then
we investigate the relation of FSLE with EKE and vorticity.
Finally, in the Conclusions we present a summary and concluding
remarks.

\section{Data and Methods}
\label{Sec:datamet}

Our dataset consists of an output from the {\it Ocean general
circulation model For the Earth Simulator} (OFES)
\cite{Masumoto2004,Masumoto2010}. This is a
near-global ocean model that has been spun up for 50 years
under climatological forcing taken from monthly mean NCEP
(United States National Centers for Environmental Prediction)
atmospheric data. After that period the OFES is forced by the
daily mean NCEP reanalysis for 48 years from 1950 to 1998. See
\cite{Masumoto2004} for additional details on the forcing.
The output of the model corresponds to daily data for the last
8 years. Horizontal angular resolution is the same in both the
zonal, $\phi$, and meridional, $\theta$, directions, with
values of $\Delta\theta =\Delta\phi=1/10^{\circ}$. The output
has been interpolated to 54 vertical z-layers and has a
temporal resolution of one day. The velocity fields
that we have used in this work correspond to the first two
years, 1990 and 1991, of the output. Vertical displacements
are unimportant during the time scales we consider here so
that, despite horizontal layers are not true isopycnals, most
fluid elements remain in their initial horizontal layer during
the time of our Lagrangian computation. Thus we use in our
analysis horizontal velocities in single horizontal layers. We
refer to recent works \cite{Ozgokmen2011,Bettencourt2012} for
Lyapunov analyses considering vertical displacements. Unless
explicitly stated, our calculations are for the second output
layer, at $7.56$ m depth, which is representative of the
surface motion but limits the effect of direct wind drag (we
have also studied the layer at $97$ m depth; results on this
layer are briefly shown in Fig. \ref{fig:timeevolution}). See
\cite{Masumoto2004} and \cite{Masumoto2010} for a
thorough evaluation of the model performance.

Among Lagrangian techniques used to quantify ocean transport
and mixing, local Lyapunov methods are being widely used. The
idea in them is to look at the dispersion of a pair of
particles as they are transported by the flow. To calculate
FTLEs, pairs of particles infinitesimally close are released
and their separation after a finite time is accounted; for
FSLEs \cite{Aurell1997} two finite distances are fixed, and
the time taken by pairs of particles to separate from the
smallest to the largest is computed. Both methods thus measure
how water parcels are stretched by the flow, and they also
quantify pair dispersion. The methods can also be tailored to
reveal two complementary pieces of information. On the one hand
they provide time-scales for dispersion and stirring process
\cite{Artale1997,Aurell1997,Buffoni1997,Lacorata2001,dOvidio2004,Haza2008,Poje2010}.
On the other, they are useful to identify Lagrangian Coherent
Structures (LCSs), persistent structures that organize the
fluid transport
\cite{Haller2000b,Haller2001,Boffetta2001,Joseph2002,Koh2002,Lapeyre2002,Haller2002,
Shadden2005,BeronVera2008,dOvidio2009,TewKai2009,Peacock2010}.
This second capability arises because the largest Lyapunov
values tend to concentrate in space along characteristic lines
which could often be identified with the manifolds (stable and
unstable) of hyperbolic trajectories
\cite{Haller2000b,Haller2001,Haller2002,Haller2011a,Shadden2005}.
Since these manifolds are material lines that can not be
crossed by fluid elements, they strongly constrain and
determine fluid motion, acting then as LCSs that organize ocean
transport on the horizontal. Thus, eddies, fronts, avenues and
barriers to transport, etc. can be conveniently located by
computing spatial Lyapunov fields. We note however that more
accurate characterization of LCSs can be done beyond Lyapunov
methods \cite{Haller2011a}, that high Lyapunov values can
correspond also to non-hyperbolic structures with high shear
\cite{dOvidio2009b}, and that an important class of LCSs is
associated to small, and not to large values of the Lyapunov
exponents \cite{Rypina2007,BeronVera2010b}.

In the present work, however, we are more interested in
obtaining the first type of information, i.e. in extracting
characteristic dispersion time-scales, quantifying the
intensity of stirring, for the different ocean regions and
seasons. In particular we want to focus on the transport
process associated to eddies and other mesoscale structures.
Previous Lagrangian analyses of the global ocean
\cite{BeronVera2008,Waugh2008} used FTLE to quantify such
horizontal stirring. This quantity depends on the integration
time during which the pair of particles is followed. FTLEs
generally decrease as this integration time increases,
approaching the asymptotic value of the infinite-time Lyapunov
exponent \cite{Waugh2008}. We find difficult to specify finite
values of this integration time for which easy-to-interpret
results would be obtained across the different ocean regions.
But for the mesoscale processes on which we want to focus,
characteristic spatial scales are related to the Rossby
Deformation Radius (RDR), with easily defined values and
latitudinal dependence (see below). Thus, we use in this paper
FSLEs as a convenient way to identify characteristics of
stirring by mesoscale processes. FSLE are also convenient in
finite ocean basins, where relevant spatial scales are also
clearly imposed \cite{Artale1997,Boffetta2000, Lacorata2001}.
As a quantifier of horizontal stirring, measuring the
stretching of water parcels, FSLEs give also information on the
intensity of horizontal mixing between water masses, although a
complete correspondence between stirring and mixing requires
the consideration of diffusivity and of the stretching
directions \cite{dOvidio2009b}.

More in detail, at a given point the FSLE (denoted by $\lambda$ in the
following) is obtained by computing the minimal time $\tau$ at
which two fluid particles, one centered on the point of study
and the other initially separated by a distance $\delta_0$,
reach a final separation distance $\delta_f$. At position
$\textbf{x}$ and time $t$, the FSLE is given by:
$\lambda (\textbf{x}, t, \delta_0, \delta_f)= \tau^{-1}
\ln(\delta_f/\delta_0)$.
To estimate the minimal time $\tau$ we
would need to integrate the trajectories of all the points
around the analyzed one and select the trajectory which
diverges the first. We can obtain a very good approximation of
$\tau$ by just considering the four trajectories defined by the
closest neighbors of the point in the regular grid of initial
conditions at which we have computed the FSLE; the spacing of
this grid is taken equal to $\delta_{0}$. The equations of
motion that describe the horizontal evolution of particle
trajectories are

\begin{eqnarray}
\frac{d\phi}{dt}&=&\frac{u(\phi, \theta, t)}{R \cos {\theta}},
\label{eqsmotion}\\
\frac{d\theta}{dt}&=&\frac{v(\phi, \theta, t)}{R},
\label{eqsmotionb}
\end{eqnarray}

where $u$ and $v$ stand for the zonal and meridional components
of the surface velocity field coming from the OFES simulations;
$R$ is the radius of the Earth ($6400$ $km$), $\phi$ is
longitude and $\theta$ latitude. Numerically we proceed by
integrating Eqs. (\ref{eqsmotion}) and (\ref{eqsmotionb}) using
a standard, fourth-order Runge-Kutta scheme, with an
integration time step $dt=6$ hours. Since information is
provided just in a discrete space-time grid, spatiotemporal
interpolation of the velocity data is required, that is
performed by bilinear interpolation. Initial conditions for
which the prescribed final separation $\delta_f$ has not been
reached after integrating all the available times in the data
set are assigned a value $\lambda=0$. A possible way to
introduce small-scale features that are not resolved by our
simulated velocity fields is by inclusion of noise terms in the
equations of motion (\ref{eqsmotionb}). We have recently shown
\cite{HernandezCarrasco2011} that the main mesoscale features
are maintained when this eddy-diffusivity is taken into
account, though sub-mesoscale structures may change
considerably. For global scales we expect the effects of noise
to be even more negligible.

The field of FSLEs thus depends on the choice of two length
scales: the initial separation $\delta_{0}$ (which coincides
with the lattice spacing of the FSLE grid and is fixed in our
computations to the model resolution,
$\delta_{0}$=$1/10^{\circ}$) and the final separation
$\delta_{f}$. As in previous works in middle latitudes
\cite{dOvidio2004,dOvidio2009,HernandezCarrasco2011}
we will focus on transport processes arising from the mesoscale
structures. In these studies $\delta_{f}$ was taken about $110
km$, which is of the order of, but larger than, the mesoscale
size in middle latitudes. Note that $\delta_f$ should be a
decreasing function of the latitude, since mesoscale structures
decrease in size with Rossby Deformation Radius (RDR). We need
not to exactly match RDR but to guarantee that our choice of
$\delta_f$ is similar but larger than mesoscale lengths, and also that
it is a smooth function to avoid inducing
artifacts. We have then chosen $\delta_f$ as
$\delta_f=1.3 |\cos\theta|$ degrees; other reasonable choices
lead to similar results to those presented here.

We compute the FSLEs by $\emph{backwards}$ time integration. In
this way we quantify the fluid deformation by {\sl past}
stirring. When computing LCSs this leads to structures easier
to interpret since they can be associated with the actual shape
of tracer filaments
\cite{Joseph2002,dOvidio2009}. However, given
that forward and backward exponents in incompressible flows are
related by temporal shifts and spatial distortions
\cite{Haller2011b}, and that we are interested in temporal and
spatial averages over relatively large scales, we do not expect
significant differences when using {\sl forward} exponents to
calculate the stirring quantifiers presented below. This was
explicitly checked in a similar framework in
\cite{dOvidio2004}.

Lagrangian measurements have been shown to correlate well with
several Eulerian quantities at several scales
\cite{Waugh2006,Waugh2008}. In particular it is pertinent to
correlate stirring with Eddy Kinetic Energy (EKE) since it is
expected that more energetic turbulent areas would also present
stronger horizontal stirring, mainly due to the spawning of
eddies (see however \cite{Rossi2008, Rossi2009}). Given an integration
period $T$ long  enough (for instance $T$= one year), the EKE
(per unit of mass) is given by: $EKE=\frac{1}{2}\left\langle
u'^{2}+v'^{2} \right\rangle$, where $u'$ and $v'$ are the
instant deviations in zonal and meridional velocities from the
average over the period $T$, and the brackets denote average
over that period. Another Eulerian measurement used in this
work is the surface relative vorticity, given by
$\omega=\frac{\partial v}{\partial x}-\frac{\partial
u}{\partial y}$, with positive (vs negative) $\omega$
associated to cyclonic (vs anticyclonic) motion in the Northern
Hemisphere (opposite signs in the Southern Hemisphere). An
additional Eulerian candidate to look for Lagrangian
correspondences is the local strain rate, but it has been shown
\cite{Waugh2006,Waugh2008} to scale linearly with $EKE^{1/2}$
and thus it will not be explicitly considered here.

Conditioned averages of $\lambda$ as a function of another variable $y$
(let $y$ be EKE$^{1/2}$ or $\omega$) introduced in
Subsection~\ref{subsec:dispersion} are obtained by discretizing the
allowed values of $y$ by binning; 100 bins were taken, each one defining
a range of values $(y_n,y_{n+1})$ and represented by the average
value $\hat{y}_n=\frac{y_n+y_{n+1}}{2}$. So, for each discretized
value of $\hat{y}_n$ the average of all the values of $\lambda$
which occur coupled with a value in $(y_n,y_{n+1})$ is computed.
The result is an estimate of the conditioned average $\tilde{\lambda}(y)$
(which is a function of $y$) at the points $\hat{y}_n$.

\section{Results}
\label{Sec:results}
\subsection{Global horizontal stirring from FSLE}

\begin{figure}
\begin{center}
\noindent\includegraphics[height=17cm, width=15cm]{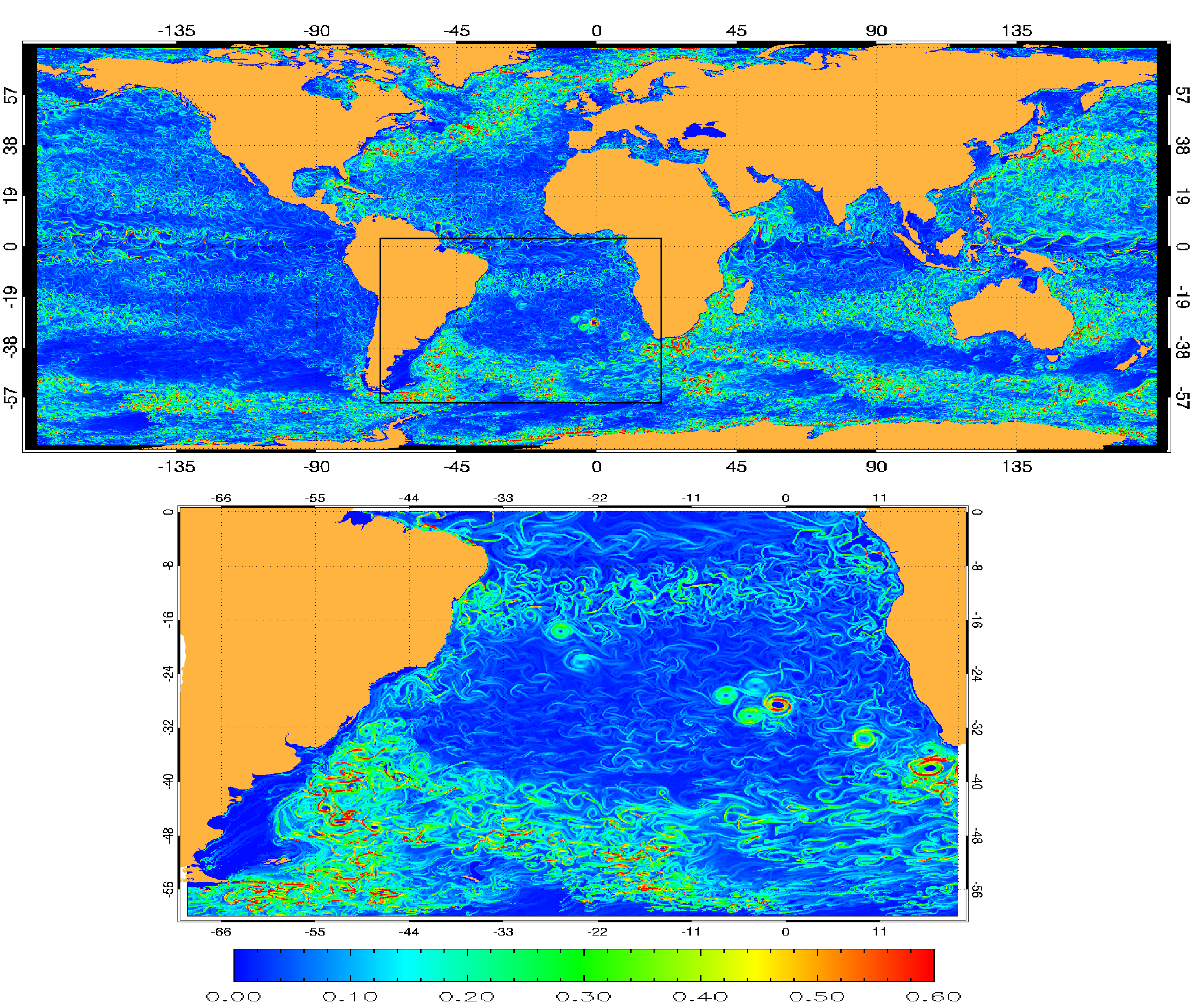}
\end{center}
\caption{Top: Snapshot of spatial distributions of FSLEs backward in
time corresponding to November 11, 1990 of the OFES output. Resolution is $\delta_0=1/10^{\circ}$.
Bottom: Zoom in the area of the box inside
top figure (South Atlantic Ocean). Coherent structures and vortices can be clearly seen.
The colorbar has units of $day^{-1}$.
}
\label{fig:instant}
\end{figure}

In Fig. \ref{fig:instant} we present a map of FSLEs at a given
time. Typical values are in the order of $0.1-0.6$ $days^{-1}$,
that correspond well to the horizontal stirring times expected
at the mesoscale, in the range of days/weeks. Spatial
structures, from filaments and mesoscale vortices to larger
ones, are clearly identified; see a representative zoom of the
South Atlantic Ocean (Bottom of Fig. \ref{fig:instant}), where
the typical filamental structures originated by the horizontal
motions are evident.

Instantaneous maps of FSLEs have a significant signature of
short-lived fast processes and are adequate to extract LCSs,
but we are more interested in slower processes at larger
scales. We have hence taken time averages of FSLEs over different
periods, in order to select the low-frequency, large-scale
signal. In this way we can easily characterize regions in the
global ocean with different horizontal stirring activity; areas
with larger values of averaged FSLEs are identified as zones
with more persistent horizontal stirring \cite{dOvidio2004},
as shown in Fig. \ref{fig:timeaverage}a. As expected, we can
observe that high stirring values correspond to Western
Boundary Currents (WBCs) and to the Antarctic Circumpolar
Current, while the rest of the ocean and the Eastern Boundary
Currents (EBCs) display significantly lower values.

\subsection{Geographical characterization of horizontal stirring}
\label{Sec:comparsion}

\begin{figure}
\begin{center}
\noindent\includegraphics[scale=0.2]{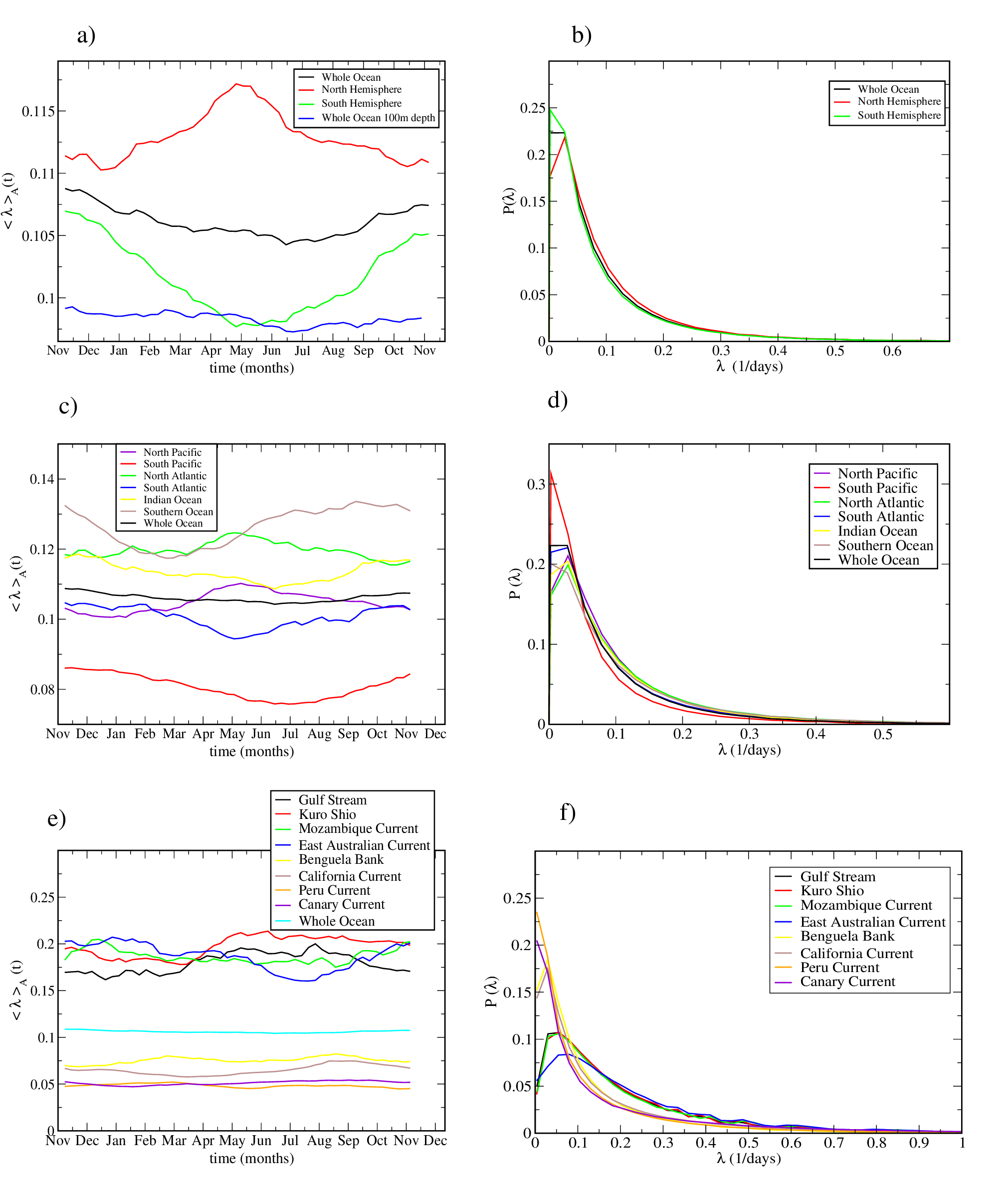}
\end{center}
\caption{Left column: Temporal evolution (from November 1st, 1990 to October 31th, 1991)
of the horizontal stirring (Spatial average of FSLEs). Right column: PDFs of the FSLEs (histograms are built from the
$\lambda$ values contained at all locations of the 52 weekly maps computed for the second simulation output year).
Top: for both hemispheres and for the whole ocean. Middle: for
different oceanic regions. Bottom: for some main currents during one simulation year. In addition to the results
from the second surface layer analyzed through the paper, panel a) shows also stirring
intensity in a layer close to 100m depth.
}
\label{fig:timeevolution}
\end{figure}

A convenient quantity used to characterize stirring in a
prescribed geographical area $A$ was introduced by
\cite{dOvidio2004}, which is simply the spatial average of the
FSLEs over that area at a given time, denoted by $<\lambda({\bf
x},t)>_{A}$. Time series of this quantity for the whole ocean
and the Northern and Southern hemispheres are shown in
Fig.~\ref{fig:timeevolution}a. It is worth noting that the
stirring intensity is typically larger in the Northern
Hemisphere than in the Southern one.

Further information can be obtained by analyzing the FSLE
Probability Distribution Functions (PDFs). In
Fig.~\ref{fig:timeevolution}b we present the PDFs for both
hemispheres and the whole ocean; the required histograms are
built using $\lambda$ values computed once every week during
one year (52 snapshots) at each point of the spatial FSLE grid
in the area of interest. Each one of these PDFs is broad and
asymmetric, with a small mode $\lambda_m$ (i.e., the value of
$\lambda$ at which the probability attains its maximum) and a
heavy tail. Similarly to what was discussed by
\cite{Waugh2006} and \cite{Waugh2008} for the FTLE case,
these PDFs are well described by Weibull distributions with
appropriate values for the defining parameters. We note that an
explicit relationship between FTLE and FSLE distributions was
derived by \cite{Tzella2010}, but we have not checked if our
flow is in the regime considered in that reference. The mode
$\lambda_m$ for the Southern Hemisphere is smaller than that of
the Northern Hemisphere. Thus, Northern Hemisphere is globally
more active in terms of horizontal dispersion than the Southern
one. The same conclusions hold when looking at seasonally
averaged instead of annually averaged quantities (not shown).

Taking into account the observed differences between Northern
and Southern Hemispheres, we have repeated the same analyses
over the main ocean basins in a search for isolating the
factors which could contribute to one or another observed
behaviors. In Fig. \ref{fig:timeevolution}c we show the time
evolution of $<\lambda>_A$ as computed over the six main ocean
basins (North Atlantic, South Atlantic, North Pacific, South
Pacific, Indian Ocean and Southern Ocean), compared to the one
obtained over the global ocean. The Southern Ocean happens to
be the most active (in terms of horizontal stirring) because of
the presence of the Antarctic Circumpolar Current, followed by
the Atlantic and Indian Oceans, and finally the Pacific. We
have also computed (Fig.~\ref{fig:timeevolution}d) PDFs of FSLE
for the different oceans. As before, we obtain broad,
asymmetric PDFs with small  modes and heavy tails. The smallest
mode $\lambda_m$ corresponds to the Southern Pacific, meaning
than there is less horizontal stirring activity in this basin,
in support of what is also visually evident in
Fig.~\ref{fig:timeevolution}c. On the opposite regime we
observe that the largest FSLE values correspond to the Southern
Ocean. For the rest of oceans the PDFs are rather coincident
with the whole ocean PDF.

We have gone further to a smaller scale, by repeating the same
analyses for the main currents in the global ocean: Gulf
Stream, Benguela, Kuroshio, Mozambique, East Australian,
California, Peru and Canary currents. As evidenced by
Fig.~\ref{fig:timeevolution}e there is a clear separation in
two groups of currents in terms of their horizontal stirring
properties: the most active currents (including Gulf Stream,
Kuroshio, Mozambique and East Australian currents, all of them
WBCs) and the least active ones (including Benguela,
California, Peru and Canary Currents, which correspond to
EBCs). The distinction remains in the PDF analysis: we can
clearly distinguish two groups of PDFs: a) narrow PDFs highly
peaked around a very small value of $\lambda$ (EBCs); b) PDFs
peaking at a slightly greater value of $\lambda$, but
significantly broader (WBCs). Since the PDFs of the WBCs are
broader, large values of FSLEs are found more frequently, i.e.,
more intense stirring occurs. This appears to be a reflection
of the well-known mechanism of Western Intensification by
\cite{Stommel1948}. Also, the asymmetry and tails of the PDFs
show that the FSLE field is inhomogeneous and that there are
regions with very different dispersion properties. Following
\cite{BeronVera2010}, asymmetry and heavy tails make the PDFs
quite different from the Gaussians expected under more
homogeneous mixing. These characteristics are then indications
that chaotic motion plays a dominant role versus turbulent,
smaller scales, dynamics. That is, the large scale velocity
features control the dynamics, something that is also reflected
in the filamentary patterns of the LCS shown in Fig.
\ref{fig:instant}.

\subsection{Seasonal characterization of horizontal stirring}

Horizontal stirring in the global ocean has a strong seasonal variability,
as shown in Fig.~\ref{fig:timeevolution}a. Maximum values of $<\lambda>_A$
in the Northern Hemisphere are reached early in that hemisphere Summer,
and minimum ones early in that hemisphere  Winter. The same happens for
the Southern hemisphere related to its Summer and Winter periods.

\begin{figure}
\begin{center}
\noindent\includegraphics[scale=0.18]{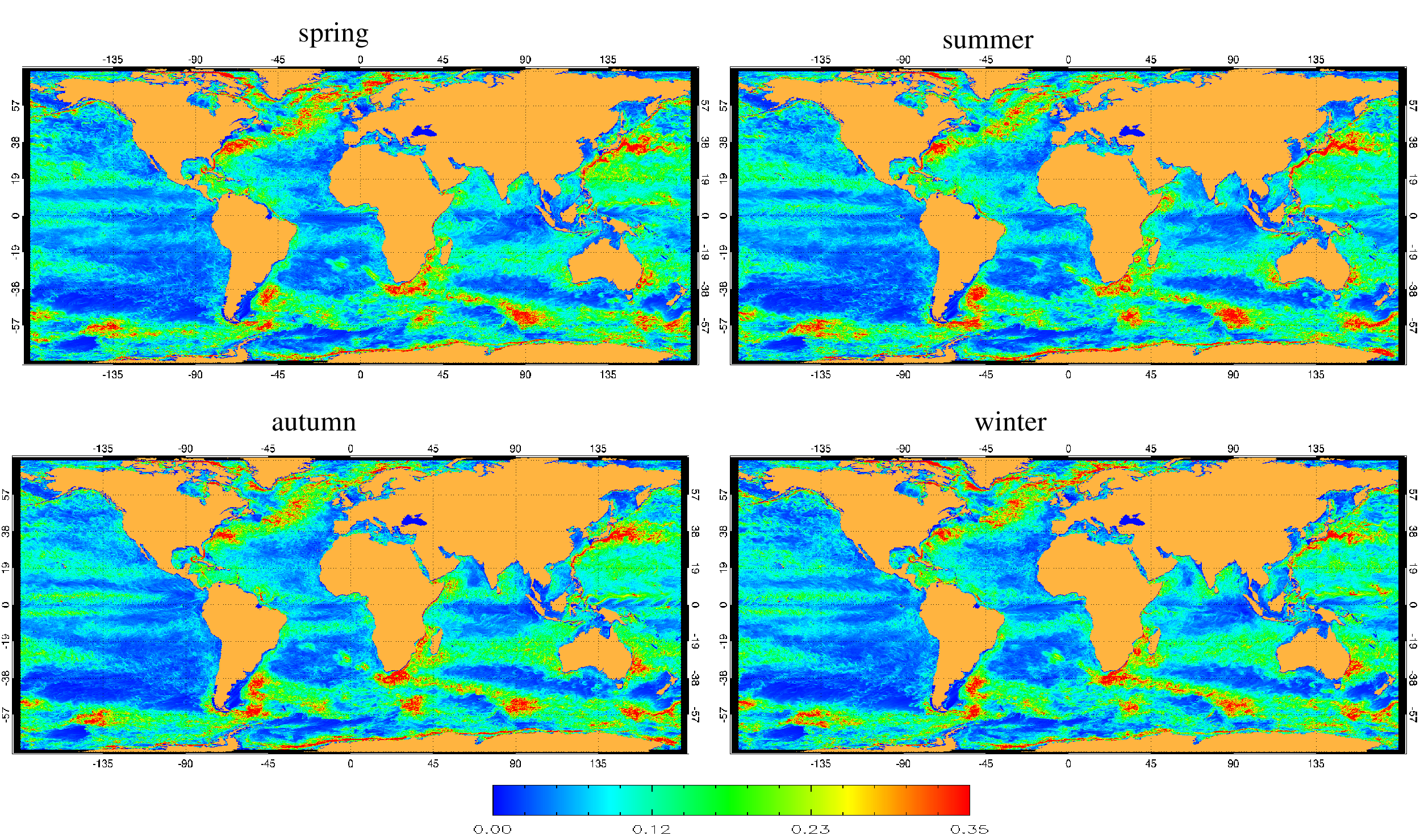}
\end{center}
\caption{Time average of the FSLEs in the Global Ocean for
the each season. Spring: from March 22 to June 22. Summer: from June 22 to September 22.
Autumn: from September 22 to December 22. Winter: from December 22 to March 22.
The colorbar has units of $day^{-1}$.
}
\label{fig:estaciones}
\end{figure}

Seasonally averaged FSLEs in the whole ocean over the four
seasons are shown in Fig.~\ref{fig:estaciones}. The spatial
pattern is rather similar in all of them, and also similar to
the annually-averaged spatial distribution shown in Fig
\ref{fig:timeaverage}a. Higher FSLE levels are found at the
Gulf Stream and Kuroshio in the Northern Hemisphere in Spring
and Summer of that hemisphere. Analogously for the  Eastern
Australia  and Mozambique Currents in the Southern Hemisphere
relative to their own Spring and Summer time.

\begin{figure}
\begin{center}
\noindent\includegraphics[scale=0.20]{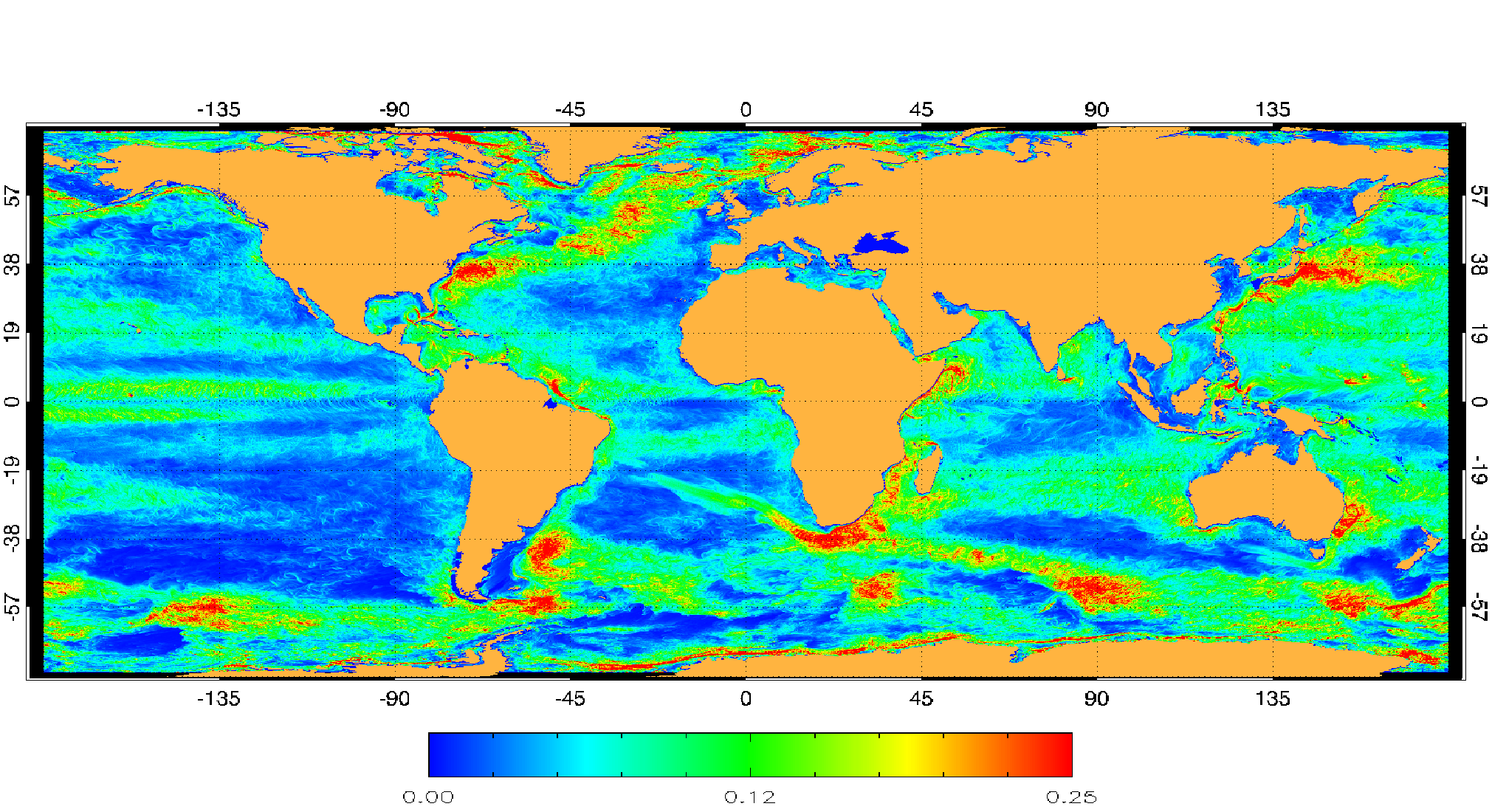}
\end{center}
\caption{Standard deviation of weekly FSLE maps of one year.
The colorbar has units of $day^{-1}$
}
\label{fig:amplitud}
\end{figure}

Following \cite{Zhai2008}, to analyze which areas are more
sensitive to seasonal changes, we computed the standard
deviation of the annual time series of FSLE (see
Fig.~\ref{fig:amplitud}). Larger values appear to correspond to
the more energetic regions thus showing a higher seasonal
variability. More information about seasonal variability of
different oceanic regions can be obtained again from
Fig.~\ref{fig:timeevolution}. Time evolution of stirring in the
North Atlantic and North Pacific, shown in
Fig.~\ref{fig:timeevolution}c, attains high values in Spring
and Summer, and minimum ones in Winter. Concerning the main
currents, we found than values of stirring in Kuroshio, Gulf
Stream, East Australia, and Mozambique currents increase in
Spring and Summer and decrease in Winter (see Fig.
\ref{fig:timeevolution}e). This seasonal variability is also
present in EBCs but the amplitude of the changes is smaller
than in WBCs.

The generic increase in mesoscale stirring in Summer time
detected here with Lyapunov methods has also been identified in
previous works and several locations
\cite{Halliwell1994,Qiu1999,Morrow2003,Qiu2004,Zhai2008} (in
most of the cases from the EKE behavior extracted from
altimetric data). Although no consensus on a single mechanisms
seems to exist (see discussion in \cite{Zhai2008}) enhanced
baroclinic instability has been proposed in particular areas
\cite{Qiu1999,Qiu2004}, as well as reduced dissipation during
Summer \cite{Zhai2008}.

We have also calculated longitudinal (zonal) averages of the
time averages of FSLE in Figs.~\ref{fig:timeaverage}a and
\ref{fig:estaciones}. This is shown in
Fig.~\ref{fig:latitud_mixing} (top figure for the Northern
hemisphere and bottom figure for the Southern one). First of
all, we see that horizontal stirring has a general tendency to
increase with latitude in both hemispheres. One may wonder if
this is a simple consequence of the decreasing value of
$\delta_f$ we take when increasing latitude. We have checked
that the same increasing tendency remains when the calculation
is redone with a constant $\delta_f$ over the whole globe (not
shown), so that this trivial effect is properly compensated by
the factor $\ln(\delta_f/\delta_0)$ in the FSLE definition, and
what we see in Fig.~\ref{fig:latitud_mixing} is really a
stronger stirring at higher latitudes. Note that this type of
dependence is more similar to the {\sl equivalent sea surface
slope variability}, $K_{sl}$, calculated from altimetry in
\cite{Stammer1997} than to the raw zonal dependency of the EKE
obtained in the same paper. Since $K_{sl}$ is intended to
represent Sea Surface Height variability with the large scale
components filtered out, we see again that our FSLE calculation
is capturing properly the mesoscale components of ocean
stirring observed by other means.

It is also clearly seen that latitudinal positions of local
maxima of stirring correspond to the main currents (e.g. Gulf
Stream and Kuroshio around 35$^{\circ}$N; Mozambique, Brazil
and East Australia around 25$^{\circ}$S). The picture in
Fig.~\ref{fig:latitud_mixing} confirms that horizontal stirring
is somehow higher in local Summer in mid-latitudes, were the
main currents are, for both hemispheres. At low and high
latitudes however the horizontal stirring is higher in local
winter-time for both hemispheres, which is particularly visible
in the Northern Hemisphere at high latitudes. A similar
behavior was noted by \cite{Zhai2008} in the subpolar North
Pacific and part of the subpolar North Atlantic for EKE derived
from altimetry. Possible causes pointed there are barotropic
instabilities or direct wind forcing.

\subsection{Lagrangian-Eulerian relations}
\label{subsec:dispersion}

Lagrangian measures such as FSLEs provide information on the
cumulative effect of flow at a given point, as it integrates
the time-evolution of water parcels arriving to that point.
They are not directly related to instantaneous measurements as
those provided by Eulerian quantities such as EKE or vorticity,
unless some kind of dynamic equilibrium or ergodicity-type
property is established so that the time-integrated effect can
be related to the instantaneous spatial pattern (for instance,
if the spatial arrangement of eddies at a given time gives an
idea about the typical time evolution of a water parcel) or
their averages. EKE gives information on the turbulent
component of the flow, which is associated to high eddy
activity, while relative vorticity $\omega$ takes into account
the shear and the rotation of the whole flow. Eventual
establishment of such dynamic equilibrium would allow to
substitute in some instances time averages along trajectories
by spatial averages, so providing a useful tool for rapid
diagnostics of sea state. Thus, we will relate the Lagrangian
stirring (as measured by the FSLEs) with an instantaneous,
Eulerian, state variable. Of course, the Lagrangian-Eulerian
relations will be useful only if the same, or only a few
functional relationships hold in different ocean regions. If
the  relation should be recalculated for every study zone, the
predictive power is completely lost.

\begin{figure}
\begin{center}
\noindent\includegraphics[scale=0.28]{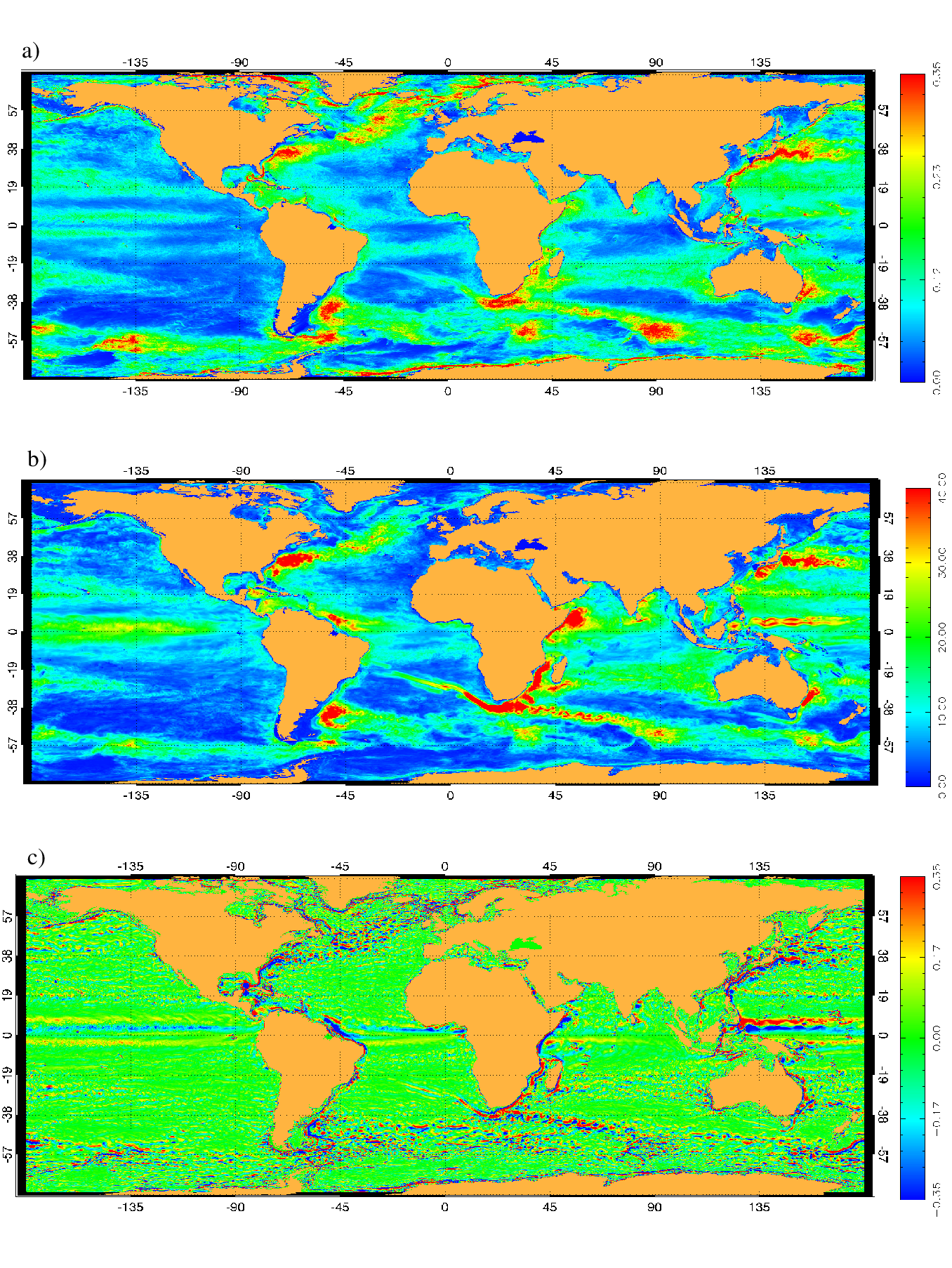}
\end{center}
\caption{a) Time average of the FSLEs in the Global Ocean.
Geographical regions of different stirring activity appear.
The colorbar has units of $day^{-1}$. b) Spatial distribution of annual $EKE^{1/2}$ (cm/s).
c) Time average of Relative Vorticity ($\omega$) in the Global Ocean. The color bar has units
of $day^{-1}$. In all the plots the averages are over the 52 weekly maps computed
 from November 1st, 1990 to October 31th, 1991.
}
\label{fig:timeaverage}
\end{figure}

\begin{figure}
\begin{center}
\noindent\includegraphics[angle=270,scale=0.15]{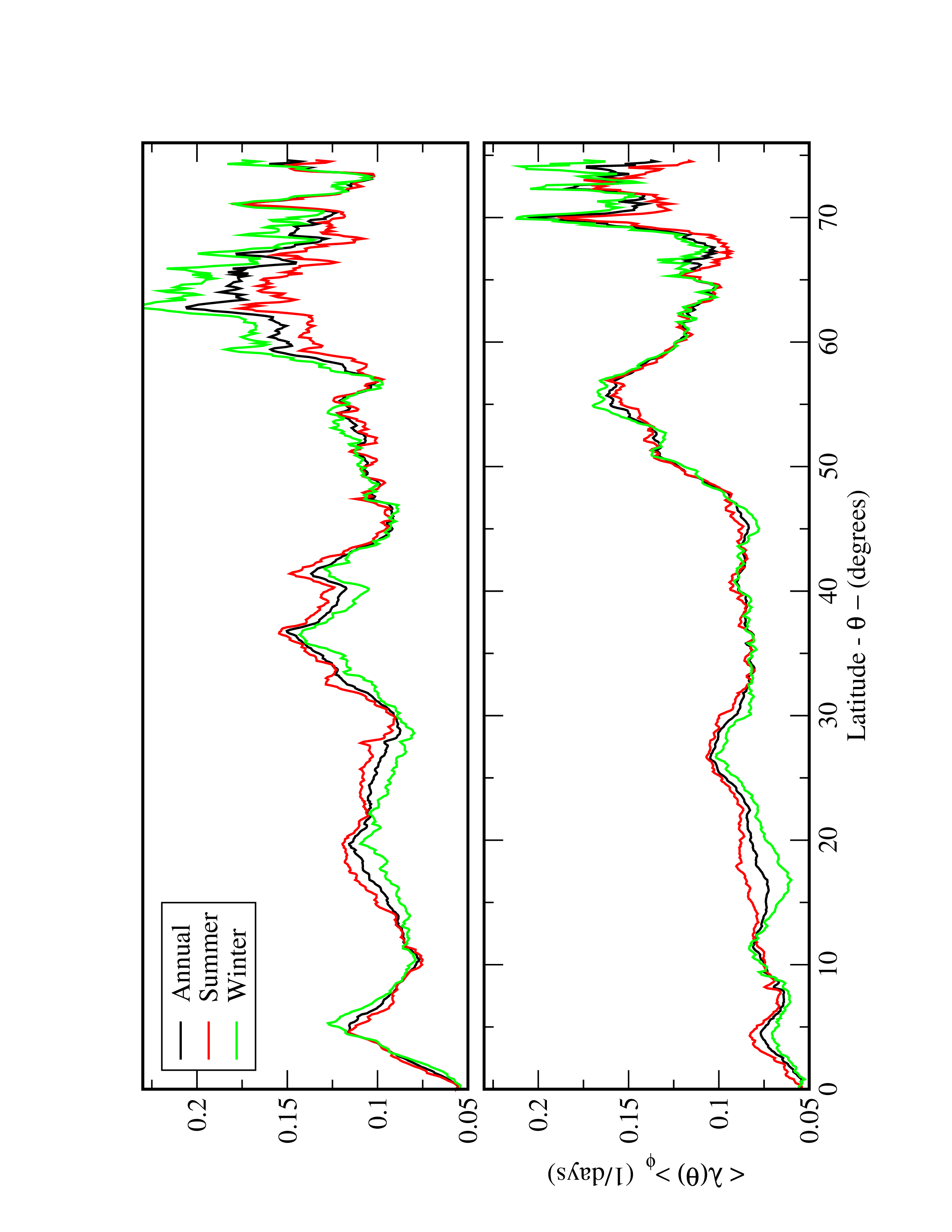}
\end{center}
\caption{Cross-ocean zonal average of the annual, relative Summer and
relative Winter time average of FSLE maps from Fig \ref{fig:timeaverage}a as
a function of latitude (expressed as absolute degrees from Equator to make
both hemispheres comparable). Top: Northern Hemisphere; bottom: Southern Hemisphere.
}
\label{fig:latitud_mixing}
\end{figure}

\begin{figure}
\begin{center}
\noindent\includegraphics[scale=0.19]{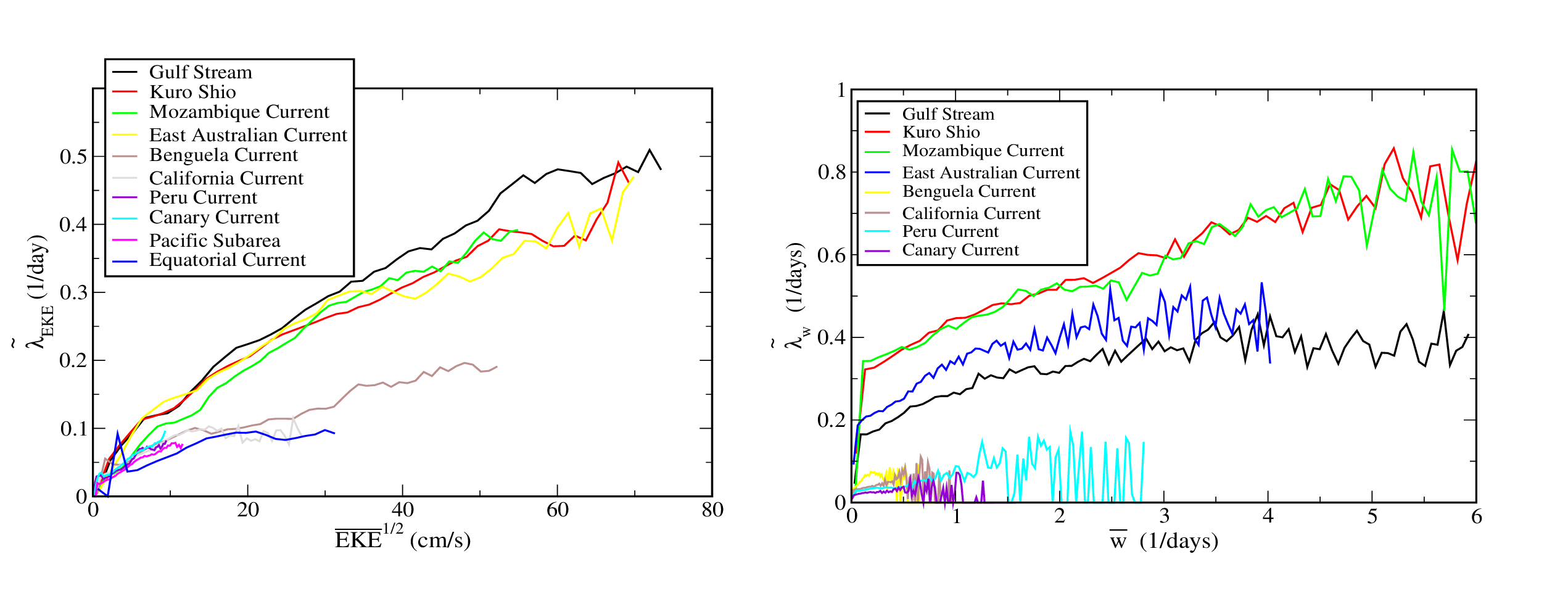}
\end{center}
\caption{Left: Lagrangian-Eulerian relations. Left: the conditional average $\tilde\lambda_{EKE}$ as
a function of its corresponding annually averaged (second year) $\overline{EKE}$ for
different regions and currents. We clearly observe two groups of relations FSLE-EKE.
Right: same plot for the conditional average $\tilde\lambda_\omega$ as a function of its corresponding
annually averaged (second year) $\overline{\omega}$. Although we observe also the same two
two groups of FSLE-$\omega$ relations, these functions are much noisier and region-dependent.
}
\label{fig:dispersion}
\end{figure}

We have thus explored the functional dependence of FLSEs with
EKE and relative vorticity. In Fig.~\ref{fig:timeaverage} the
time average of these three fields is shown. Comparing FSLEs
(Fig.~\ref{fig:timeaverage}a) and EKE
(Fig.~\ref{fig:timeaverage}b), we see that high and low values
of these two quantities are generally localized in the same
regions. There are a few exceptions, such as the North Pacific
Subtropical Countercurrent, which despite being energetic
\cite{Qiu1999} does not seem to produce enough pair dispersion
and stretching at the scales we are considering. It was already
shown by \cite{Waugh2006} and \cite{Waugh2008} that
variations in horizontal stirring are closely related to
variations in mesoscale activity as measured by EKE. Note the
similarity, with also an analogous range of values, of the EKE
plot in Fig. ~\ref{fig:timeaverage}b), obtained from a
numerical model, to that of \cite{Waugh2008} (first figure),
which is obtained from altimetry data. In \cite{Waugh2006} a
proportionality between the stretching rate (as measured by
FTLE) and $EKE^{1/4}$ was inferred for the Tasman Sea (a
relation was found but no fit was attempted in the global data
set described in \cite{Waugh2008}). In order to verify if a
similar functional dependence between FSLE and EKE could hold
for our global scale dataset, we have computed different
conditioned averages (see Section \ref{Sec:datamet}), shown in
Fig.~\ref{fig:dispersion}: in the left panel we present the
conditioned average $\tilde\lambda(EKE)$, while in the right
panel $\tilde\lambda(\omega)$ is shown; both functions were
derived from the time averaged variables shown in
Figure~\ref{fig:timeaverage}.

\begin{figure}
\begin{center}
\noindent\includegraphics[scale=0.15]{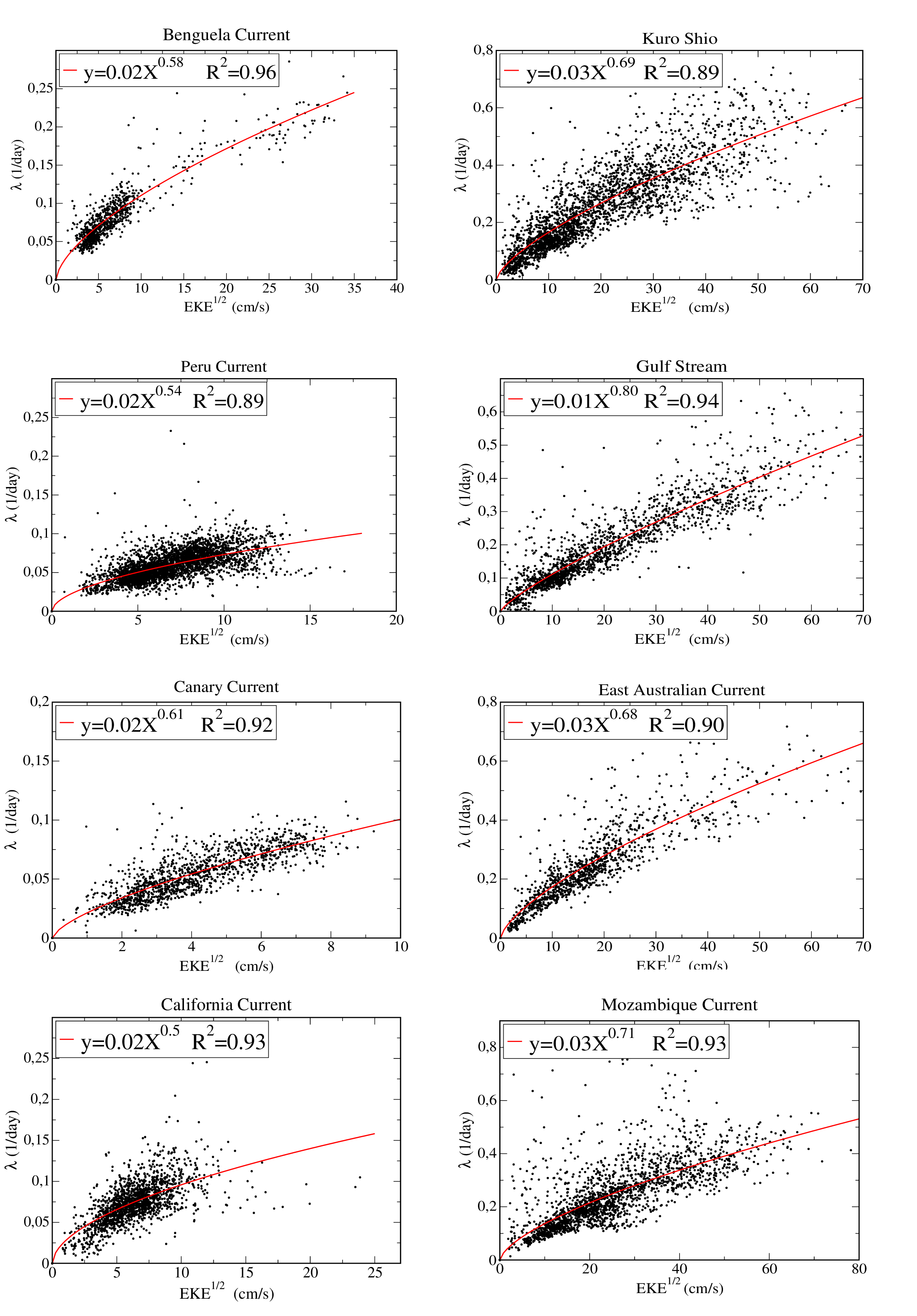}
\end{center}
\caption{Scatter plots showing temporally averaged FSLE values at different spatial points in regions
of Fig.~\ref{fig:timeaverage}a, and EKE values (as displayed in Fig.~\ref{fig:timeaverage}b)
at the same points. The regions displayed here are eight
of the main currents. Fittings of the type $y=c X^b$ are also displayed,
where $y$ is the temporal mean of FSLE and $X$ is $\textrm{EKE}^{1/2}$.  Note that this implies
$<\textrm{FSLE}>=c ~\textrm{EKE}^\alpha$ with $\alpha=b/2$.
}
\label{fig:dispersionscat}
\end{figure}

The smooth curve depicted in Fig.~\ref{fig:dispersion}, left,
is an indication of a well-defined functional relationship
between $\overline{\lambda}$ and $\overline{EKE}$, similar to
the ones found by \cite{Waugh2006} and \cite{Waugh2008} from
altimeter data. Notice however that the plot just gives
conditioned averages, but the conditioned standard deviation
-which is a measure of randomness and fluctuations- is not
negligible. An idea of the scatter is given for selected areas
in Fig.~\ref{fig:dispersionscat}. Considerably less compact
relationships were obtained in the Mediterranean sea
\cite{dOvidio2009}. Fig.~\ref{fig:dispersionscat}
shows that very different dynamical regimes identified by
different values of $\lambda$ may correspond to the same level
of EKE. As a Lagrangian diagnostic, we believe that FSLE is
more suitable to link turbulence properties to tracer dynamics
than Eulerian quantifiers such as EKE. FSLEs provide
complementary information since very energetic areas, with
large typical velocities, do not necessarily correspond to high
stretching regions. A paradigmatic example is a jet, or a shear
flow, where small dispersions may be found because of the
absence of chaotic trajectories. A functional relation between
$\overline{\lambda}$ and $\overline{\omega}$ is also obtained
(Fig.~\ref{fig:dispersion}, right), although it is much noisier
and probably worse-behaved. When particularizing for the
different regions, we see that for EKE the WBCs are all roughly
associated with one particular functional relation for the
conditioned average $\overline{\lambda}$ while EBCs gather
around a different one. None of the two prototype
Lagrangian-Eulerian relations fits well to the relation
$\lambda \propto EKE^{1/4}$ proposed for FTLE by
\cite{Waugh2006} from altimeter data in the Tasman sea. Data
are too scarce to make a reliable fitting for the conditioned
average, in particular for the EBC. In
Fig.~\ref{fig:dispersionscat} we see that  relations of the
form $\lambda \propto EKE^{\alpha}$ could be reasonably fitted
to scatter plots of the data, with $\alpha$ larger than the
$0.25$ obtained in \cite{Waugh2006}, specially for WBC were
$\alpha$ is in the range $(0.34,0.40)$. This quantitative
difference of our results with \cite{Waugh2006} may rest upon
the fact that they considered just the Tasman Sea and we
consider the different oceans. Other sources for the difference
could be that we are using FSLE of velocity data from a
numerical model, instead of FTLE from altimetry, or that they
use a grid of relatively low resolution $0.5^{\circ} \times
0.5^{\circ}$, while ours is $0.1^{\circ} \times 0.1^{\circ}$.
Maybe their coarser resolution is not enough to resolve
filaments which are the most relevant structures in our FSLE
calculations. Despite this the qualitative shape of the
Lagrangian-Eulerian relations is similar to the previous works
\cite{Waugh2006,Waugh2008}.

In order to analyze the ocean regions beyond boundary currents,
we have also computed the conditioned averages for the
Equatorial Current and for a $40^{\circ}$ longitude by
$20^{\circ}$ latitude sub-region centered at $245^{\circ}$
longitude and $-30^{\circ}$ latitude in the middle of the
sub-tropical gyre in the Pacific Ocean (and hence an area of
scarce horizontal stirring activity). We see
(Fig.~\ref{fig:dispersion}, left) that the EBC
Lagrangian-Eulerian relation is valid for these two areas. We
have also verified that the relations derived from
annually-averaged quantities remain the same for seasonal
averages (not shown). The important point here is the
occurrence of just two different shapes for the EKE-FSLE
relations across very different ocean regions, which may make
useful this type of parametrization of a Lagrangian quantity in
terms of an Eulerian one. For the relations of FSLE in terms of
relative vorticity, a distinction between WBC and EBC still
exists but the results are less clear and class separation is
not as sharp as in the case of EKE (see
Fig.~\ref{fig:dispersion}, right). For instance, Gulf Stream
and Kuroshio, despite being both WBC, do not seem to share the
same Lagrangian-Eulerian relation, which limits its usefulness.

\section{Conclusions}
\label{Sec:conclusion} In this paper we have studied the space
and time variability of horizontal stirring in the global ocean
by means of FSLE analysis of the outputs of a numerical model.
Similarly to what has been done in previous works, FSLEs can be
taken as indicators of horizontal stirring. Being Lagrangian
variables, they integrate the evolution of water parcels and
thus they are not completely local quantities. We have taken
averages to analyze two main time scales (annual and seasonal)
and three space scales (planetary scale, ocean scale and
horizontal boundary scale). Our velocity data were obtained by
using atmospheric forcing from NCEP. Structures and dynamics at
small scales will be probably more realistic if forcing with
higher resolution observed winds, as in \cite{Sazaki2006}. But
since we have not studied the first model layer which is
directly driven by wind, and we have focused on averages at
relatively large time and spatial scales, we do not expect much
differences if using more detailed forcing.

Horizontal stirring intensity tends to increase with latitude,
probably as a result of having higher planetary vorticity and
stronger wind action at high latitudes, or rather, as argued in
\cite{Zhai2008} because of barotropic instabilities. Certainly, new
studies are required to evaluate these hypothesis.
At a planetary scale we observe a significantly different behavior
in the Northern hemisphere with respect to the Southern
Hemisphere, the first being on average more active in terms of
horizontal stirring than the second one. This difference can
probably be explained by the greater relative areas of
subtropical gyres in the Southern Hemisphere with small
stirring activity inside them, which compensates in the
averages the great activity of the Antartic Circumpolar
Current. At an ocean scale, we observe that the level of
stirring activity tends to decay as the size of subtropical
gyres increases, what is an indication that the most intense
horizontal stirring takes place at the geographical boundaries
of ocean basins. For that reason, we have finally analyzed the
behavior of stirring at boundary scale, which is mainly related
to WBCs and EBCs. EBCs behave in a similar way to ocean
interior in terms of all the quantities we have computed,
including the Lagrangian-Eulerian relations. Thus, the main hot
spots of horizontal stirring in the ocean are WBC. The observed
small mode in the global FSLE PDFs also indicates that
horizontal stirring is not very intense for the vast majority
of the ocean, but the heavy tails indicate the existence of
large excursions at some specific, stretched locations (e.g.,
inside the WBCs and other smaller scale currents active enough
to generate stirring). This type of uneven distribution is
characteristic of multifractal systems arising from large scale
chaotic advection, something that was discussed for oceanic
FSLEs in \cite{HernandezCarrasco2011}.

Regarding seasonal variability, generally we observe stronger
stirring during each hemisphere's Summer time. Medium and high
latitudes behave however in the opposite way: stirring is more
active during the hemisphere Summer for medium latitudes and
during the hemisphere Winter for high latitudes. Medium
latitudes are strongly affected by the behavior of WBC, which
experience intensification of horizontal stirring during Summer
\cite{Halliwell1994,Qiu1999,Morrow2003,Qiu2004,Zhai2008}. As
commented before, high latitude Winter intense stirring could
be the result of a stronger action of wind during that period
or of barotropic instabilities \cite{Zhai2008}, and dedicated
studies are required to evaluate these hypothesis.

Finally, we have studied the connection between time-extended
Lagrangian FSLEs and instant Eulerian quantities such as EKE
and relative vorticity. For the case of EKE, the different
ocean regions give rise to just two different Lagrangian-Eulerian
relations, associated to an intense or a weak stirring regimes.
The existence of these two regimes implies that pair dispersion
and stretching strength are larger in a class of ocean areas
(represented by WBC) than in another (e.g. EBC) at mesoscales,
even when having the same EKE.


%
%
%
%
%
%

%
%
%
%

\section*{Acknowledgments}
I.H.-C., C.L. and E.H.-G. acknowledge support from MICINN and
FEDER through project FISICOS (FIS200760327); A. Turiel has received
support from Interreg TOSCA project (G-MED09-425) and Spanish MICINN
project MIDAS-6 (AYA2010-22062-C05-01). The OFES simulation was
conducted on the Earth Simulator under the support of JAMSTEC.
We thank Earth Simulator Center-JAMSTECH team
for providing these data.

%
%
%
%
%


\end{document}